\documentclass[floatfix,prl,aps,twocolumn,showpacs,preprintnumbers,amsmath,amssymb]{revtex4}

\usepackage{graphicx}
\usepackage{dcolumn}
\usepackage{bm}

\begin{document}
\title{Radio Frequency Selective Addressing of Localized Particles in a Periodic Potential}

\author{H. Ott}
\author{E. de Mirandes}
\author{F. Ferlaino}
\author{G. Roati}
\author{V. T\"urck}
\author{G. Modugno}
\author{M. Inguscio}

\affiliation{LENS and Dipartimento di Fisica, Universit\`a di
Firenze, and INFM, Via Nello Carrara 1, 50019 Sesto Fiorentino,
Italy}

\date{\today}
\begin{abstract}
We study the localization and addressability of ultra cold atoms
in a combined parabolic and periodic potential. Such a potential
supports the existence of localized stationary states and we show
that using a radio frequency field allows to selectively address
the atoms in these states. This method is used to measure the
energy and momentum distribution of the atoms in the localized
states. We also discuss possible extensions of this scheme to
address and manipulate particles in single lattice sites.
\end{abstract}

\maketitle Periodic potentials have been used with great success
in a series of experiments with ultra cold atoms
\cite{Salomon1996,Raizen1996,Cataliotti2001,Greiner2002}. For a
trapped sample the periodic potential is usually accompanied with
an additional parabolic confinement
\cite{Cataliotti2001,Greiner2002}. In most experiments involving
trapped Bose-Einstein condensates only the properties of the
ground state, which are not changed dramatically by the parabolic
confinement, are important. However, for the excited states of the
system the additional harmonic confinement must be taken into
account. This is especially true for atomic Fermi gases, where the
Pauli principle enforces a population of higher energetic states.
The qualitative different nature of the single particle states can
be seen from recent experiments with ultra cold fermions which
have found evidence for a localization of the particles in such a
potential \cite{Modugno2003,Ott2003,Pezze2004}. In contrast to the
Mott insulating phase \cite{Greiner2002}, this localization
process is independent from the interaction and is a pure
consequence of the potential shape. It is therefore important to
understand its properties and physical consequences, nonetheless
because most of the theoretical work on quantum phase transitions
involving fermions
\cite{Rigol2003a,Albus2003,Büchler2003,Lewenstein2003} are based
on pure homogeneous systems. A combined periodic and parabolic
potential is also interesting for possible applications in quantum
information and was recently proposed for the implementation of a
qubit register for fermions \cite{Viverit2004}. A possible
addressability of individual atoms in single lattice sites is
thereby an intriguing vision.

In this work we study ultra cold atoms in a parabolic magnetic
potential which is superimposed in its weak direction with a one
dimensional optical lattice. The combined potential possesses two
distinct classes of eigenstates, which --- depending on their
energy --- either extend symmetrically around the trap center or
are localized on the sides of the potential. We use a radio
frequency technique to address the particles in localized states
and to measure the density of state along the lattice direction as
well as the momentum distribution of the particles. Due to the
localization of the particles, the radio frequency field addresses
the particles in a defined spatial region and we discuss the
possibility of extending this scheme to manipulate individual
atoms in single lattice sites.

In the experiment we prepare an ultra cold cloud of $^{87}$Rb in
the combined potential by forced evaporative cooling. The magnetic
trapping potential has an axial and radial oscillation frequency
of $\omega_{\textrm a}=2\pi\times16$\,Hz and $\omega_{\textrm
r}=2\pi \times 197$\,Hz and the optical lattice
($\lambda=830$\,nm) can be adjusted between $0<s<10$, where $s$
measures the potential height in units of the recoil energy
$E_r=h^2/2m\lambda^2$. The atoms are prepared in the
spin-polarized $\mid F=2,m_{\textrm F}=2\rangle$ state and the
temperature is between 500 and 600 nK. Here we are only interested
in the single particle behavior; in this frame bosons and fermions
show the same features, as long as interparticle interactions can
be neglected.

\begin{figure}
\includegraphics[width=8cm]{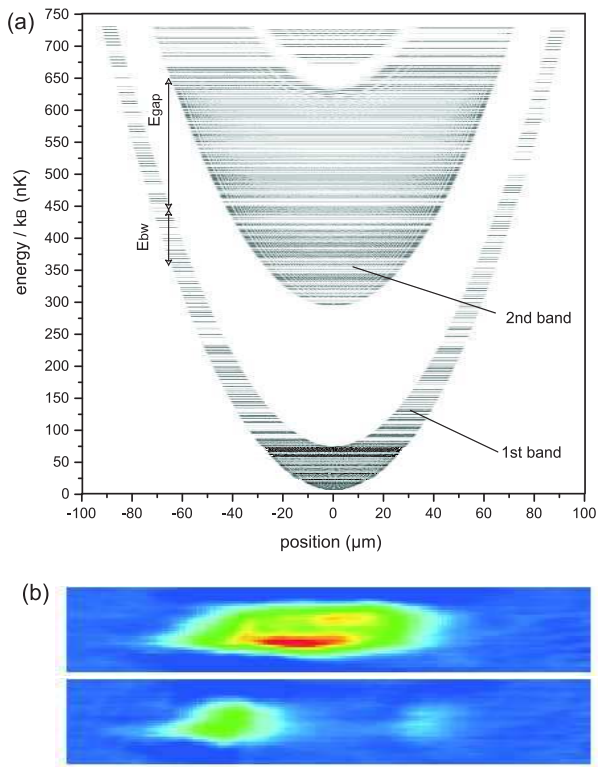}
\caption{\label{figure1}(a) Spectrum of the Hamiltonian:
representation of the 1D spectrum of the single particle
Schr\"odinger equation for a combined parabolic and periodic
potential with $s$=3. Each line represents one eigenstate of the
system, which is plotted as density profile in grayscale. The
vertical position of the profile corresponds to the energy of the
eigenstate. (b) Cloud of atoms without rf field (upper image) and
with rf field (lower image) after 1.5 ms time of flight.}
\end{figure}

To understand the properties of the combined potential we first
solve the 1D Schr\"odinger equation in the direction of the
lattice,
\begin{equation}
\label{seq} \left[-\frac{\hbar^2}{2m}\frac{\partial^2}{\partial
x^2}+\frac{1}{2}m\omega ^2x^2+\frac{s}{2}E_r(1-cos4\pi
x/\lambda)\right]\psi=E_n\psi,
\end{equation}
where $E_n$ is the energy of the $n$th eigenstate. This
Hamiltonian has also recently been studied in tight binding
approximation \cite{Rigol2003,Hooley2003}. In Fig.\,1a we show a
density plot of the first 1000 eigenfunctions of (\ref{seq}). Each
line in Fig.\,\ref{figure1}a corresponds to a density plot of the
wave function. For low energies we find delocalized states that
spread symmetrically around the potential minimum. Above a
threshold energy, the wave functions of the eigenstates become
localized on both sides of the potential. If we look at higher
energies, a second group of eigenstates appears, centered again
around the trap minimum. It is straightforward to identify this
shell-like structure with the well known band picture for a pure
periodic potential. This becomes particularly clear if one looks
at the accessible energy values for a given position which
correspond to the calculated bandwidth $E_{\textrm{bw}}$ and band
gap $E_{\textrm{gap}}$ of a pure sinusoidal potential, only
shifted in energy by the local value of the parabolic potential.
As a direct consequence, the system does not have an absolute but
a spatially varying energy gap. The temperature of the cloud is
chosen that the bandwidth of the first band is much smaller than
the average energy of the atoms, thus providing a high population
of the localized states. In the two radial directions, the
particles occupy the harmonic oscillator states.

We now introduce our experimental technique to prepare and address
the particles in the localized states. After the end of the
evaporation a radio frequency (rf) field is applied in order to
induce spin flip transitions. Because of the magnetic radial
confinement this removes all atoms whose wavefunction has a
spatial overlap with the magnetic field shell where the resonance
condition $h\nu=\mu_{\textrm B}B(\mathbf{r})/2$ is fulfilled.
Periodically sweeping (1\,kHz rate) the radio frequency within an
interval $\Delta\nu=\nu_{\textrm{up}}-\nu_{\textrm{low}}$ we
define a spatial region in which the atoms are removed from the
potential. After 100\,ms of rf field we image the atoms which are
left in the potential. In Fig.\,\ref{figure1}b we show an
absorption image of the atomic cloud after $1.5$\,ms time of
flight without and with rf field. In the latter case we remain
with two clouds, located at the edges of the original cloud. Atoms
in these two clouds are trapped in localized states and the result
directly shows, that the rf field is capable to address the atoms
in a defined spatial region. We have checked that (i) without the
optical lattice all atoms are removed from the trapping potential
and (ii) after switching off the rf field the atoms remain on
their position in the trap. Only for small lattice heights we can
observe a slow motion of the two clouds towards the trap center.
This is due to presence of collisions which allow the bosons to
hop between different localized states \cite{Ott2003}. For a
spin-polarized Fermi gas this effect is absent.

\begin{figure}
\includegraphics[width=8cm]{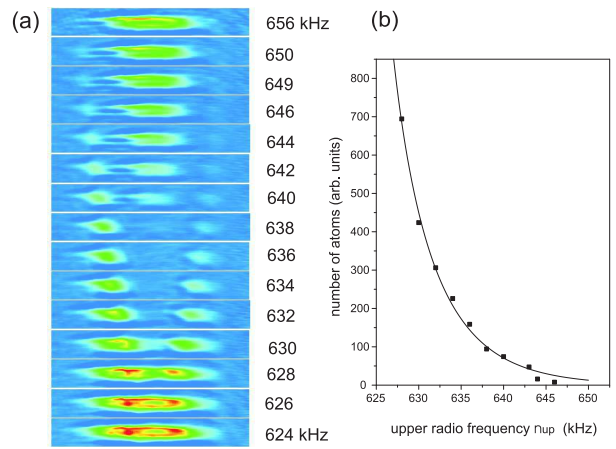}
 \caption{\label{figure2}Energy distribution of
localized states in a lattice with $s$=9. (a) Scan of the rf field
through the cloud. The indicated frequencies are the upper
frequency $\nu_{\textrm{up}}$ of the rf field, the sweep amplitude
of the field is 3\,kHz. (b) Atom number in the left cloud in
dependence on $\nu_{\textrm{up}}$. The straight line is a fit with
an axial density of states that is proportional to $E^{-1/2}$ (see
text).}
\end{figure}

In our recent work \cite{Modugno2003,Ott2003,Pezze2004} we were
able to detect the localization of the atoms by looking at the
center of mass position of the whole cloud. Here, we can employ
this kind of rf spectroscopy to look at the energy distribution of
the atoms in the localized states. In Fig.\,\ref{figure2}b we show
a series of absorption images where we have scanned the rf field
fixing the frequency interval $\Delta\nu=3$\,kHz. Increasing the
rf frequency we start to remove atoms from the center of the trap.
The hole in the spatial distribution deepens, until the lower
frequency bound is higher than the resonance frequency at the trap
bottom: atoms in the center are no longer removed from the
potential and we observe three clouds until for even higher
frequencies the displaced peaks disappear and the cloud is again
unaffected by the rf field. In Fig.\,\ref{figure2}c we show the
number of atoms in the left cloud in dependence on upper frequency
of the rf field. Due to the localization these atoms have an axial
energy which is higher than
$E_{\textrm{up}}=2h(\nu_{\textrm{up}-\nu_0})$, where $\nu_0$ is
the resonance frequency in the trap center. Thus, the number of
atoms in the cloud is determined by the density of states in the
axial direction: $N\propto\int_{E_{\textrm
up}}^{\infty}n(E_{\textrm a})\rho_{\textrm a}(E_{\textrm
a})dE_{\textrm a}$, where $\rho_{\textrm a}(E_{\textrm a})$ is the
density of states in the axial direction and $n(E_{\textrm
a})=e^{-E_{\textrm a}/k_{\textrm B}T}$ is the axial energy
distribution. As shown in Ref.\,\cite{Hooley2003} the density of
states for energies larger than the bandwidth of the first band is
predicted to be proportional to $E^{-1/2}$. Using this value we
fit our data leaving the temperature and a constant of
proportionality as free parameters \cite{temperature}. The result
which is shown in Fig.\,\ref{figure2}b is consistent with our
data.

\begin{figure}
\includegraphics[width=8cm]{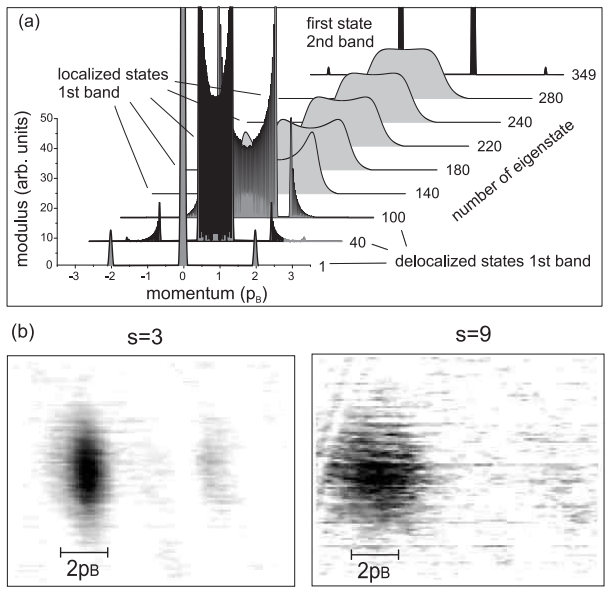}
\caption{\label{figure3} Momentum distribution of the localized
states. (a) Fourier transform of selected eigenstates of the
spectrum shown in Fig.\,\ref{figure1}a. The first three states are
delocalized and lie within the first band, the last state is the
lowest state of the second band. The states in between are
localized states. (b) Absorption image of a cloud of atoms such as
shown in Fig.\,\ref{figure2}b after 10\,ms time of flight for
$s=3$ and $s=9$.}
\end{figure}

We now turn to the momentum distribution of the localized states.
For a potential depth of $s=3$ we have calculated the Fourier
transform of the wave functions of the eigenstates.
Fig.\,\ref{figure3}a shows the momentum distribution for selected
eigenstates within the first and second band. The lowest
eigenstate shows the well known peak distribution at multiples of
twice the Bragg momentum. With increasing energy these peaks
broaden and develop a substructure. The states of our interest are
the localized states, whose distribution spreads over the first
Brillouin zone ($\pm p_{\textrm B}$) and is similar for all
states, regardless the energy of the state. In the radial
direction, the momentum distribution is determined by the
temperature of the cloud. Consequently, localized clouds as shown
in Fig.\,\ref{figure2} are expected to exhibit an anisotropic
expansion. In Fig.\,\ref{figure3}b we show an absorption image of
a localized cloud after 10\,ms time of flight for $s=3$ and $s=9$.
The measured aspect ratio of $2.5$ for $s=3$ is highly anisotropic
and directly proves the nonclassical momentum distribution of the
localized states. For $s=9$ the cloud expands much faster in the
direction along the lattice (horizontal direction). Indeed we
calculate a 2 times larger momentum distribution for $s=9$ with
respect to $s=3$ which leads to a nearly isotropic expansion.

\begin{figure}
\includegraphics[width=8cm]{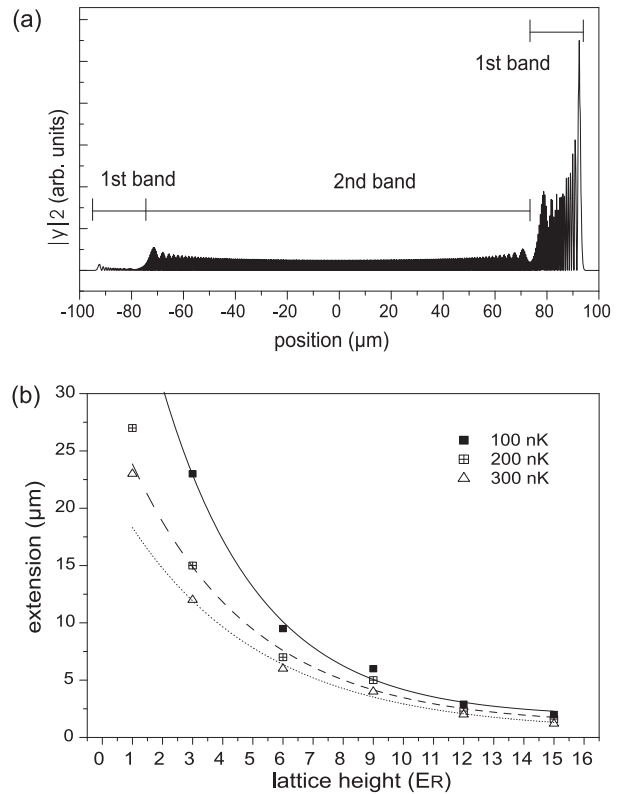}
\caption{\label{figure4}(a) Tunnelling between the bands: density
distribution of the 594th eigenstate for a potential with $s=0.3$.
(b) Extension of the localized states in the first band in
dependence on the lattice height for three different energies
($k_{\mathrm{B}}\times 100$ nK, $k_{\mathrm{B}}\times 200$ nK, and
$k_{\mathrm{B}}\times 300$ nK,).}
\end{figure}

We finish this work with discussing the localization process and
its possible applications. For small lattice heights, the
localization of the states is prevented by inter band transitions
and as an example we show in Fig.\,\ref{figure4}a an excited state
for $s=0.3$. The wave function exhibits substantial contributions
from both bands and a particle in this state is no longer
localized. On the other side the extension of the localized state
shrinks with increasing lattice height (Fig.\,\ref{figure4}b) and
the smallest possible extension is given by the ground state in
each lattice site. For our parameters we find that for $s=30$ the
eigenstates are mainly located within a single lattice site. This
result is of particular interest because it shows that for similar
experimental conditions like in the Mott insulator experiment
\cite{Greiner2002} a localization of the particles within one
lattice site is possible without a repulsive interaction. Indeed,
Viverit et al.\,\cite{Viverit2004} have shown that if an atomic
Fermi gas is loaded in a combined parabolic and periodic potential
even an occupancy with exactly one atom per lattice site can be
achieved. Another intriguing consequence of the localization is
the addressability of single lattice sites: the potential gradient
discriminates the resonance condition for an atomic transition in
each lattice site if the transition depends on the external
potential. In our setup the magnetic potential leads to a
spatially varying Zeeman splitting within the
$|F=2\rangle$--manyfold and thus a very weak radio frequency
should allow -- in principle -- for the manipulation of the atoms
within one lattice site. To get a reasonable discrimination and a
sufficiently high Rabi frequency, the resonance condition between
adjacent lattice sites should be shifted by something about 10 kHz
which would require a gradient of 300 G/cm \cite{gradient}.
Thereby a linear potential is more favorable than a parabolic one
where the frequency shift is changing along the lattice. For well
defined experimental conditions it would be also desirable to have
an optical confinement in the radial direction because otherwise
particles with lower axial but higher radial energy might also be
resonant with the radio frequency.

In conclusion we have proved that particles in a combined periodic
and parabolic potential are trapped in localized states. We have
used a radio frequency field to induce spatially resolved
transitions in order to remove the particles from the potential.
Doing so it was possible to measure the axial density of states
and the momentum distribution of the localized states. The
experiment has implications in various directions. First it shows
that an inhomogeneous periodic potential exhibit a qualitatively
different phenomenology compared to a homogeneous system. Second
the experiment directly evidences a new localization mechanism
which is independent from the interaction. Especially for atomic
Fermi gases such a potential can be interesting to create single
lattice occupancy \cite{Viverit2004}. Third we show that in
principle the particles can be addressed spatially resolved. For
future experiments it should be possible to extend the scheme to
the manipulation of particles in single lattice sites which would
constitute a major progress in "quantum engineering" with ultra
cold atoms.

This work was supported by MIUR, by EU under Contract. No.
HPRICT1999-00111, and by INFM, PRA "Photonmatter". H.\,O. and
V.\,T. were supported by EU with Marie Curie Individual
Fellowships, under Contracts No. HPMF-CT-2002-01958 and
HPMF-CT-2002-01649.


\begin{references}
\bibitem{temperature}
The x-axis in Fig.\,\ref{figure2}b is not directly the energy in
the potential because our trap bottom exhibits long term drifts.
As the different data points were taken in time order a linear
drift leads to a scaling of the energy axis which can be
compensated by leaving the temperature in the fitting procedure as
a free parameter.
\bibitem{tilt}
The asymmetry in the population of the two clouds rises from a
small horizontal tilt of our magnetic trap with respect to
gravity. As a consequence, the potential minimum does not coincide
with the magnetic field minimum.
\bibitem{uce}
We cannot directly calculate $E_{\textrm{up}}$ from the radio
frequency because the bottom of our magnetic trapping potential is
drifting with time and therefore $\nu_0$ is unknown, but we can
take the position of the cloud as a reference for the energy.
\bibitem{gradient} For this gradient, the required lattice height
for a localization within one lattice site and a simultaneous
suppression of Zener tunnelling is $s>10$.
\end{references}
\end{document}